\documentclass[12pt,epsf]{article}
\usepackage{amsmath}
\usepackage{amssymb}
\usepackage{epsfig}
\usepackage{graphicx}
\begin{document}
\def\a{\alpha}
\def\b{\beta}
\def\c{\varepsilon}
\def\d{\delta}
\def\e{\epsilon}
\def\f{\phi}
\def\g{\gamma}
\def\h{\theta}
\def\k{\kappa}
\def\l{\lambda}
\def\m{\mu}
\def\n{\nu}
\def\p{\psi}
\def\q{\partial}
\def\r{\rho}
\def\s{\sigma}
\def\t{\tau}
\def\u{\upsilon}
\def\v{\varphi}
\def\w{\omega}
\def\x{\xi}
\def\y{\eta}
\def\z{\zeta}
\def\D{{\mit \Delta}}
\def\G{\Gamma}
\def\H{\Theta}
\def\L{\Lambda}
\def\F{\Phi}
\def\P{\Psi}

\def\S{\Sigma}

\def\o{\over}
\def\beq{\begin{eqnarray}}
\def\eeq{\end{eqnarray}}
\newcommand{\gsim}{ \mathop{}_{\textstyle \sim}^{\textstyle >} }
\newcommand{\lsim}{ \mathop{}_{\textstyle \sim}^{\textstyle <} }
\newcommand{\vev}[1]{ \left\langle {#1} \right\rangle }
\newcommand{\bra}[1]{ \langle {#1} | }
\newcommand{\ket}[1]{ | {#1} \rangle }
\newcommand{\EV}{ {\rm eV} }
\newcommand{\KEV}{ {\rm keV} }
\newcommand{\MEV}{ {\rm MeV} }
\newcommand{\GEV}{ {\rm GeV} }
\newcommand{\TEV}{ {\rm TeV} }
\def\slash#1{\ooalign{\hfil/\hfil\crcr$#1$}}
\def\diag{\mathop{\rm diag}\nolimits}
\def\Spin{\mathop{\rm Spin}}
\def\SO{\mathop{\rm SO}}
\def\O{\mathop{\rm O}}
\def\SU{\mathop{\rm SU}}
\def\U{\mathop{\rm U}}
\def\Sp{\mathop{\rm Sp}}
\def\SL{\mathop{\rm SL}}
\def\tr{\mathop{\rm tr}}

\baselineskip 0.7cm

\begin{titlepage}

\begin{flushright}
UCB-PTH-10/14
\end{flushright}

\vskip 1.35cm
\begin{center}
{\Large \bf Eternal Inflation with Liouville Cosmology}

\vspace*{0.7cm}

\begin{center}
{Yu Nakayama}
\end{center}
\vspace*{-0.2cm}
\begin{center}
{\it Berkeley Center for Theoretical Physics, \\ 
University of California, Berkeley, CA 94720, USA 
}
\vspace{3.8cm}
\end{center}

\abstract{We present a concrete holographic realization of the eternal inflation in $(1+1)$ dimensional Liouville gravity by applying the philosophy of the  FRW/CFT correspondence proposed by Freivogel, Sekino, Susskind and Yeh (FSSY). The dual boundary theory is nothing but the old matrix model describing the two-dimensional Liouville gravity coupled with minimal model matter fields. In Liouville gravity, the flat Minkowski space or even the AdS space will decay into the dS space, which is in stark contrast with higher dimensional theories, but the spirit of the FSSY conjecture applies with only minimal modification. We investigate the classical geometry as well as some correlation functions to support our claim. We also study an analytic continuation to the time-like Liouville theory to discuss possible applications in $(1+3)$ dimensional cosmology along with the original FSSY conjecture, where the boundary theory involves the time-like Liouville theory. We show that the decay rate in the $(1+3)$ dimension is more suppressed due to the quantum gravity correction of the boundary theory.}
\end{center}
\end{titlepage} 

\setcounter{page}{2}

\section{Introduction}
The concept of metastable states in quantum systems has long been studied since the earliest days of the quantum mechanics as important quantum phenomena associated with classically forbidden tunnelling effects. In the context of the quantum field theory, the most convenient approach to tackle the problem is to use bounce instanton of the Euclidean action \cite{Kobzarev:1974cp}\cite{Coleman:1977py}. The bounce instanton solution enables us to compute a semiclassical decay rate of the metastable states in the quantum field theory. The subsequent expansion of the bubble can be understood as an analytic continuation of the bounce instanton solution.

In discussing the quantum cosmology, in particular eternal inflation, the effects of the quantum gravity might be important. These effects on the vacuum decay of the universe has been largely unknown partly due to the lack of well-defined off-shell formulation of the quantum gravity. At the level of the semiclassical Einstein gravity, the Coleman-De Luccia (CDL) instanton \cite{Coleman:1980aw} leads to a classical evolution of the universe after decay through the bubble nucleation. Will the effects of the quantum gravity modify the semiclassical analysis or not? Certainly, when the decay rate is very large, and the size of the bubble is as small as the Planck scale, the ultra-violet properties of the quantum gravity should be important. Also for a very large scale, the effect of cosmological horizon might be relevant. 

Furthermore, one may even argue against the possibility of quantum tunnelling transition in quantum gravity. In general relativity, there is no strict notion of ``higher" energy states and ``lower energy" states because the total Hamiltonian is zero irrespective of the states we consider due to the Hamiltonian constraint. Of course, even without the gravity, the energy conservation demands that the quantum tunnelling is actually microscopically reversible, and it is only because the final state has a larger entropy that the quantum decay looks irreversible. The entropy of the universe with the quantum gravity contribution, however, is yet to be discussed. In addition, as of 2010, there is no experimental evidence that the universe has shown quantum phase transition from our ancestor vacua in the context of eternal inflation.\footnote{Arguably, the tininess of the cosmological constant of our universe may be the observed evidence that the (string) landscape exists. However, logically speaking, even if the tininess of the cosmological constant is a consequence of the landscape and anthropic argument, it is not immediately inevitable to assume that the quantum tunnelling transition must have occurred.}

It would be of great importance, therefore, that we establish the possibility of the quantum tunnelling with possible quantum gravity modifications and its theoretical consequence in the solvable quantum theory of gravity, where the lack of renormalizability, ill-defined Euclidean path integral, unsatisfactory off-shell formulation of string theory etc do not impede our theoretical analysis. In this paper, for this purpose, we study the quantum vacuum transition in $(1+1)$ dimensional Liouville gravity, which is one of the simplest but highly non-trivial examples of solvable quantum gravity.

In the Euclidean two-dimensional gravity, the quantum gravity corrections to the vacuum decay was first studied in \cite{Zamolodchikov:2006xs}. There, the analogue of the CDL instanton in Liouville gravity was studied in the semiclassical limit, and they showed that the amplitudes are consistent with the exact result from the matrix model computation. In this paper, we study the Lorentzian continuation of their results and the subsequent dynamics of vacuum tunnelling in $(1+1)$ dimensional Liouville cosmology.

In our analysis on $(1+1)$ dimensional cosmology with possible ``eternal inflation", we show a concrete example of the boundary holographic description in the spirit of Freivogel-Sekino-Susskind-Yeh (FSSY) conjecture \cite{Freivogel:2006xu}\cite{Bousso:2008as}\cite{Sekino:2009kv} known as Friedman-Robertson-Walker cosmology / conformal field theory (FRW/CFT) correspondence. They proposed a dual field theory framework to understand the dynamics of bubble nucleation and eternal inflation in $(1+3)$ dimensional Einstein gravity. In $(1+3)$ dimension, the details of the boundary theory have been largely unknown and it is very difficult to make any concrete prediction or check the validity of their proposal.  We propose that in the $(1+1)$ dimensional Liouville gravity, the dual theory is nothing but the old matrix model for $c<1$ matter \cite{Brezin:1990rb}\cite{Douglas:1989ve}\cite{Gross:1989vs} (with possible gauging of the trivial scaling factor). We show that the philosophy of FRW/CFT correspondence is completely in agreement with the old matrix model. Consequently, our proposal not only provides the concrete realization of the FSSY philosophy, but also introduces a new perspective on the old matrix model.

We will also study the implication of the Liouville bounce instanton in the time-like Liouville theory by further analytic continuation. The time-like Liouville theory  appears as a two-dimensional quantum gravity with large (positive) central charge from the matter sector. They are ubiquitous in the dual formulation of the higher dimensional cosmology: it appears in the $(1+3)$ dimensional original FSSY conjecture, dS/dS duality \cite{Alishahiha:2004md}\cite{Dong:2010pm} for $(1+2)$ dimensional cosmology and so on. We identify the decay rate of the Liouville gravity with the decay rate of the higher dimensional universe. The quantum gravity corrections to the two-dimensional system is now interpreted as quantum gravity corrections to the $(1+3)$ dimensional universe. As we will discuss in this paper, a surprising consequence of the identification is the modification of the decaying rate of the universe over a significantly long period. After a suitable analytic continuation of the decay rate of the two-dimensional Liouville theory, we see the emergence of the bound for the bounce instanton of the effective $(1+3)$ dimensional field theory. Extremely large instanton action, which would give rise to an extremely long life-time of the universe beyond the Poincar\'e recurrence time of our ancestor universe, is now forbidden. 

The organization of the paper is as follows. In section 2, we first review the Euclidean bounce instanton in the two-dimensional Liouville gravity, and study its analytic continuation to $(1+1)$ dimensional cosmology. In section 3, we compute some correlation functions and propose the holographic dual description by using the old matrix model. In section 4, we further study the effect of bounce instanton in time-like Liouville theory and its consequence in the FSSY conjecture for $(1+3)$ dimensional cosmology. We conclude with further discussions in section 5. In appendix A, we show the most generic bounce instanton solution in two-dimensional Liouville theory with arbitrary cosmological constant and evaluate its classical action.

\section{Instanton in Liouville gravity}
In this paper, we would like to study the dynamics of quantum tunnelling phenomena in $(1+1)$ dimensional quantum gravity. Its Euclidean analogue is the two-dimensional quantum gravity with a Euclidean thermal nucleation process inside a metastable (classical) two-dimensional system \cite{Zamolodchikov:2006xs}. The aim of this paper is to establish the precise connection between the Euclidean field theory computation and the real-time Lorentzian process.

Without the quantum gravity effect, the classical Euclidean thermal nucleation process is related to the quantum tunnelling via the simple Wick rotation of the underlying field theory. For concreteness, let us consider the $(1+1)$ dimensional landscape of vacua described by a single Landau-Ginzburg field $X(t,x)$  which will be Wick rotated to a two-dimensional Euclidean field $X(x_1,x_2)$, with the potential $V(X)$. Each vacua correspond to the extremum of the potential $V'(X) = 0$. At each extremal point $X_i$, one can expand the potential
\begin{align}
V(X) \simeq V(X_i) + a_2(X_i-X)^2 + a_3(X_i-X)^3 + a_4(X_i-X)^4 \cdots \ .
\end{align}

The low energy effective field theory at each conformal fixed point is governed by the degeneracy of the potential around the extrema. For instance, when $a_2 \neq 0$, it is given by the massive $c=0$ (trivial) theory, and when $a_2 = a_3 =0$ and $a_4\neq 0$, it is described by the $c=1/2$ theory that has the same universality class with the critical Ising model (or a single free fermion). More precisely, the vacuum whose effective potential is given by $V(X) = X^{2(k-1)}$ around the extremum is described by the unitary minimal model with the central charge
\begin{align}
c_k = 1-\frac{6}{k(k+1)} \ .
\end{align}

Now we would like to study the bubble nucleation in this setup.
 Let us consider a two-dimensional (Euclidean) flat space filled with a metastable vacuum corresponding to an extremum of the Landau-Ginzburg potential $V(X)$. Since it is metastable, there is a non-zero probability of quantum tunnelling into a lower energy vacuum. The energy difference between the ``false" vacuum and the ``true" vacuum is denoted by $\mu (>0)$. In the above example, $\mu$ dictates the potential difference of the Landau-Ginzburg field: $\mu = V(X_i) - V(X_j)$.
The surface tension\footnote{The introduction of the surface tension necessarily breaks the conformal symmetry of the matter sector in UV. In the above Landau-Ginzburg description, the tension could be computed by using the semiclassical WKB method \cite{Coleman:1977py}: $\sigma = \int_{X_i}^{X_j} dX\sqrt{2V(X)}$.} of the domain wall is denoted by $\sigma$, and the vacuum decay rate can be computed as 
\begin{eqnarray}
P_{\mathrm{classical}} \sim \exp \left(-\frac{\pi\sigma^2}{\mu}\right) \ \label{twa}
\end{eqnarray}
without any quantum gravity corrections.

The argument in the exponent is nothing but the Boltzmann factor of the two-dimensional Euclidean field theory, and one can study the quantum tunnelling rate between the  corresponding $(1+1)$ dimensional vacua by simple analytic continuation. The tunnelling rate is again given by the same formula \eqref{twa} and once a quantum bubble is nucleated, it expands rapidly almost at the speed of light: the Euclidean instanton solution $X_E(x_1,x_2)$ will be Wick rotated to an expanding bubble solution $X_L(t,x_2) = X_E(x_1=-it,x_2)$.

We would like to study the quantum gravity correction to the $(1+1)$ dimensional quantum tunnelling process. Classically, $(1+1)$ dimensional gravity is trivial: the Einstein-Hilbert action is topological and there is no dynamical degrees of freedom. The metric equation only gives a Virasoro-like constraint for the matter energy momentum tensor: $T_{\mu\nu} = \lambda g_{\mu\nu}$, where $\lambda$ is an effective cosmological constant. Quantum mechanically, due to the anomaly of the path integral measure, the Liouville mode becomes dynamical. 
In the conformal gauge, therefore, the quantum gravity in $(1+1)$ dimension is described by the Liouville field theory \cite{David:1988hj}\cite{Distler:1988jt} (see \cite{Nakayama:2004vk} for a review).
The action of the Liouville field theory is given by
\begin{align}
S_L = \int d^2x \left( \frac{1}{4\pi b^2}\partial_a \varphi \partial^a \varphi + \mu e^{2\varphi} \right) \ . \label{lact}
\end{align}
The Liouville field theory is a conformal field theory and the central charge is given by $c = 1+6(b+b^{-1})^2$. Although the quantum gravity in $(1+1)$ dimension is always strongly coupled, the Liouville field theory by itself has a classical limit $b\to 0$, where the semiclassical analysis is valid ($b^2$ is regarded as $\hbar$ in quantum mechanics).

In the subsequent sections, we will investigate instanton solutions of the Euclidean Liouville theory and their analytic continuation to the Lorentzian theory. The discussion goes in parallel with the higher dimensional study by \cite{Coleman:1980aw}. Before the detailed analysis, however, we would like to point out one fundamental difference that is particular to $(1+1)$ dimension. In higher dimensions, the positive energy (i.e. more suppressing Boltzmann weight) indicates that the space-time is described by de Sitter (dS) space while the negative energy indicates that it is described by Anti de Sitter (AdS) space. In the Liouville gravity, although it sounds counterintuitive at first, the  relation is precisely {\it opposite}. The positive energy implies that the space time is described by the AdS space while the negative energy implies that it is described by the dS space. This is due to the fact that the conventional Liouville equation that is designed for bounded Boltzmann weight obtained from \eqref{lact}:
\begin{align}
\partial \bar{\partial} \phi = \pi b^2 \mu e^{2\phi} 
\end{align}
with positive $\mu$ describes the negatively curved hyperbolic space in the Euclidean signature (i.e. AdS space after continuation to the Lorentzian signature).

As a result, a natural decaying process in $(1+1)$ dimensional Liouville gravity is from the AdS space to the dS space. In the Euclidean field theory,  \cite{Zamolodchikov:2006xs} indeed showed that the bubble in the flat Euclidean space is described by a punctured sphere with positive curvature. One of the key issues in the $(1+1)$ dimensional cosmology with quantum tunnelling is to understand this counterintuitive decaying process from the space-time viewpoint as we will see.

\subsection{Euclidean instanton}
We first give a brief review of  effects of the two-dimensional quantum gravity in a nucleation process inside a metastable two-dimensional system \cite{Zamolodchikov:2006xs}. As in \cite{Zamolodchikov:2006xs}, we study the decay of the metastable vacuum with zero energy to a (more) stable vacuum with the negative energy. In the flat space region, the Liouville action takes
\begin{eqnarray}
S_{\mathrm{Liouville}} = \frac{1}{4\pi b^2} \int d^2x \partial_a\varphi \partial^a \varphi \ , \label{unl}
\end{eqnarray}
where $b$ is related to the central charge of the Liouville theory as $c_{\mathrm{Liouville}} = 1 + 6(b + b^{-1})^2$. Note that the Liouville cosmological constant in front of $e^{2\varphi}$ is zero in the flat space. As a quantum gravity, the total central charge should balance between the Liouville sector and the matter sector: $c_{\mathrm{Liouville}} + c_{\mathrm{matter}} = 26$. The classical limit of the Liouville quantum gravity is achieved in $b\to 0$ limit, where the matter central charge should take a large negative value.\footnote{To achieve this limit in the context of the Landau-Ginzburg theory introduced in the previous section, we can incorporate spectator ghost fields to accommodate the large negative central charge.} Obviously, the Liouville field takes a constant value (say $\varphi =0$) outside of the bubble.

The nucleation of the true vacuum will give a ``wrong sign" Liouville cosmological constant. In this process, the Liouville gravity obtains a following additional contribution to the action from the droplet (bounce instanton):
\begin{eqnarray}
S_{\mathrm{Liouville}}^{\mathrm{bounce}} = \int_{\mathrm{droplet}} d^2x \left(\frac{1}{4\pi b^2} \partial_a\varphi \partial^a \varphi - \mu e^{2\varphi} \right) \ . 
\end{eqnarray}
Note that the Liouville cosmological constant $(-\mu)$ here has an opposite signature compared with a conventional Liouville action. Inside the bubble, the Liouville equation is solved by a sphere metric:
\begin{align}
e^{2\varphi} = \frac{4R^2a^2}{(1+a^2z\bar{z})^2} \ ,
\end{align}
where $R^2 = \frac{1}{4\pi b^2\mu}$.
For later purposes of analytic continuation, let us rewrite the metric in the polar coordinate $z = e^{X+i\theta}$ and express it in another conformal form:
\begin{align}
ds^2 = e^{2\varphi} dzd\bar{z} = a(X)^2 (dX^2 + d\theta^2) \ .
\end{align} 
The warp factor $a(X)$ is given by
\begin{align}
a(X) = a_0 e^{X} \ (\text{for} \ X > X_0)  \ , \ \ a(X) = \frac{2R a e^{X}}{{1+a^2 e^{2X}}} \ (\text{for} \ X<X_0) \ ,
\end{align}
where $X= X_0$ is the location of the domain wall in the thin wall approximation. We call the solution as Coleman-De Luccia-Zamolodchikov-Zamolodchikov (CDLZZ) instanton.

In the semiclassical regime, where $b \ll 1$, the contribution of this bounce instanton in the Liouville gravity has been computed in \cite{Zamolodchikov:2006xs}, which gives a quantum gravity corrected formula for the vacuum decay rate:
\begin{eqnarray}
P_{\mathrm{Liouville}} \sim \left(1+ \frac{\pi\sigma^2}{b^{-2}\mu}\right)^{-b^{-2}} \ . \label{zz}
\end{eqnarray}
It is easy to see that, in the semiclassical limit $b \to 0$ while keeping $\mu$ fixed, the result reduces to the classical result \eqref{twa}, where the quantum gravity corrections vanish: $\lim_{b\to 0} P_{\mathrm{Liouville}} \sim \exp \left(-\frac{\pi\sigma^2}{\mu}\right) \sim P_{\mathrm{classical}}$. On the other hand, in the weak metastability limit $\mu \to 0$ while keeping $b$ fixed, the vacuum decay has only power-like suppression: $\lim_{\mu \to 0} P_{\mathrm{Liouville}} \sim (b^{-2}\mu/\pi\sigma^2)^{b^{-2}}$. The decay rate of the false vacuum is, therefore, enhanced due to the effect of the two-dimensional Liouville gravity.

The result above is based on the semiclassical Liouville theory analysis which is valid in the small $b$ expansion. However, we have some evidence that the result is even true for finite $b$ \cite{Zamolodchikov:2006xs}. For instance, it is known that the partition function of the two-dimensional quantum gravity on a sphere with a puncture shows an exact scaling behaviour \cite{Knizhnik:1988ak}: $Z \sim \mu^{b^{-2}}$. The sphere with a puncture can be seen as a quantum analogue of the classical bounce instanton configuration in the weak metastability limit $\mu \to 0$. We see that this partition function completely agrees with the behavior of \eqref{zz} in the weak metastability limit.

One can generalize the computation of the decay rate between vacua with generic cosmological constants. We present the detailed derivation and summarize the results in Appendix A. It is imperative to notice that the decay rate is always power-like for finite $b$. The exponential classical result is obtained by taking $b\to 0$ limit.

\subsection{Lorentzian continuation}
We would like to study the Lorentzian continuation of the bounce instanton solution to study the ``eternal inflation" in $(1+1)$ dimensional Liouville gravity. As in the Euclidean geometry, the $(1+1)$ dimensional Liouville gravity has a counterintuitive feature that vacua with negative ($\simeq$ more stable) cosmological constant are the dS space and vacua with positive ($\simeq$ more unstable) cosmological constant are the AdS space. However, as we will see, this counterintuitive picture by itself does not affect the causal structure of the Lorentzian continuation of the bounce instanton very much. 

\begin{figure}[htbp]
    \begin{center}
    \includegraphics[width=0.3\linewidth,keepaspectratio,clip]
      {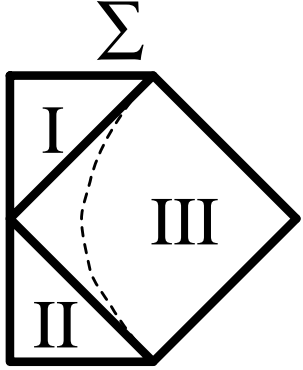}
    \end{center}
    \caption{ The Penrose diagram for the maximal extension of the Lorenzian continuation of CDLZZ instanton. The broken line denotes the position of the domain wall: $X=X_0$.}
    \label{sign}
\end{figure}

For definiteness, we begin with the tunnelling from a vacuum with a zero cosmological constant to another vacuum with a negative cosmological constant. The Euclidean geometry
\begin{align}
ds^2 = a(X)^2 \left(dX^2 + d\theta^2 \right) 
\end{align}
can be analytically continued to the Lorentzian space by 
\begin{align}
X \to X \ ,  \ \ \theta \to i T \ .
\end{align}
The continued geometry describes region III of the Penrose diagram of the full solution shown in fig 1. In the thin wall approximation studied in the previous subsection, $X= X_0$ is the location of the domain wall. More precisely, the warp factor is given by 
\begin{align}
a(X) = a_0 e^{X} \ (\text{for} \ X > X_0)  \ , \ \ a(X) = \frac{2R a e^{X}}{{1+a^2 e^{2X}}} \ (\text{for} \ X<X_0) \  
\end{align}
with the corresponding metric
\begin{align}
ds^2 = a(X)^2 \left(dX^2 - dT^2 \right) \ .
\end{align}

In region I, one can perform a different analytic continuation
\begin{align}
X \to T+ i\pi \ , \  \ \theta  \to iY \ 
\end{align}
to obtain an analogue of FRW universe in higher dimension.
The metric is given by
\begin{align}
ds^2 = a(T)^2 \left(-dT^2 + dY^2 \right) 
\end{align}
with the warp factor
\begin{align}
a(T) = \frac{R^2e^{T}}{{1-e^{2T}}} 
\end{align}
which describes the $(1+1)$-dimensional ``open de-Sitter space". One can extend the coordinate in the past as well, and the maximally extended conformal diagram is presented in fig 1. 

The geometry has a $U(1)$ isometry that can be compared with the $SO(2,1)$ isometry of the $(3+1)$-dimensional CDL background. The $U(1)$ symmetry is time-like in region I and space-like in region III. The situation is similar to the Schwartzshild black hole where the ``energy" becomes space-like across the horizon.

More generically, we can study the quantum tunnelling between vacua with non-zero cosmological constants. One can perform the same analytic continuation to the metric 
\begin{align}
ds^2 = a(X)^2 \left(dX^2 + d\theta^2 \right) 
\end{align}
to obtain the Lorentzian geometry with $U(1)$ isometry. For instance, the quantum tunnelling between vacua with negative cosmological constant (i.e. dS space to dS space) has the conformal diagram presented in fig 2. It is interesting to observe that in contrast to higher dimensions, the terminating vacuum is always dS space in $(1+1)$ dimension, if any, so in our discussions there seems no problem associated with the AdS crunch that might invalidate the holographic approach to the eternal inflation.

\begin{figure}[htbp]
    \begin{center}
    \includegraphics[width=0.3\linewidth,keepaspectratio,clip]
      {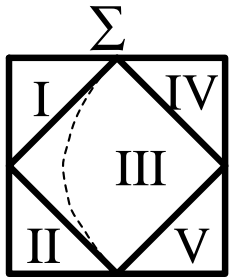}
    \end{center}
    \caption{ The Penrose diagram for the decay between the dS vacua. The broken line denotes the position of the domain wall: $X=X_0$.}
    \label{sign}
\end{figure}
\section{Boundary Theory --- FSSY conjecture and matrix model}
FSSY proposal \cite{Freivogel:2006xu} is a novel holographic way to describe our (1+3) dimensional universe from the viewpoint of the cosmic census taker \cite{Susskind:2007pv} sitting at a future infinity of the stable universe via FRW/CFT correspondence. Physics of the cosmic census taker is described by a boundary theory living at the boundary of the constant conformal time of the FRW universe. Alternatively, the FSSY proposal can be seen as a light-like Kaluza-Klein reduction of the ``conformal boundary theory" which is $(1+2)$ dimension. The conformal boundary is light-like, which is observed from the Census Taker at time-like infinity, and the theory living at the light-like boundary is further reduced to the one living on the two-sphere, which coincides with the boundary of the constant conformal time. 

In our Liouville gravity, the boundary is nothing but a point,\footnote{The boundary of the de-Sitter space in (1+1) dimension consists of two distinct points. The Penrose diagram fig 1 and 2 describe one boundary attached to the domain wall. The other boundary is represented as spatial infinity (in region I) and it could be cut off by the other domain wall by studying the two-fold cover of the diagram. We focus on the point that coincides with one of the domain walls in the future boundary described as the broken lines in the Penrose diagram. As long as we discuss the analytic continuation of the CDLZZ instanton in the thin wall approximation, the two points do not communicate with each other.} 
and the boundary theory is expected to possess a finite number of degrees of freedom. A small difference from the original FSSY proposal is the conformal boundary (in region I in Figure 1). Here the boundary is space-like and there is no obvious notion of the Census Taker here, but the philosophy and the computational detail of the Kaluza-Klein reduction goes in paralell with the original FSSY proposal as we will see.

In section 3.1, we first compute the correlation function of the boundary theory by using the proposal made in \cite{Freivogel:2006xu} with some modified interpretation. In section 3.2, we propose the dual boundary theory to be the old matrix model \cite{Brezin:1990rb}\cite{Douglas:1989ve}\cite{Gross:1989vs} proposed for $c<1$ non-critical string theory.
\subsection{FSSY correlation function}
To understand the nature of the boundary theory, we would like to compute the boundary correlation functions from the scalar field propagating in the CDLZZ instanton background. For simplicity, we assume that the background  Liouville field theory is semiclassical, and the scalar mode $\chi$ is minimally coupled with the Liouville gravity only through the kinetic term. As a direct consequence, the scalar mode is automatically massless: otherwise it must show further couplings with the Liouville field.

The Lorentzian correlation function requires a specification of the boundary condition, and we use a similar prescription proposed by FSSY in the higher dimensional CDL instanton background \cite{Freivogel:2006xu}. For this purpose, we first compute the Euclidean two-point function propagating in the two-dimensional geometry
\begin{align}
ds^2 = a(X)^2 \left( dX^2 + d\theta^2 \right) \ .
\end{align}
The warp factor behaves as $a(X) \sim e^{X}$ near $X\to \infty$ and $a(X) \sim \frac{e^{X}}{{1+e^{2X}}}$ near $X\to - \infty$. The $\theta$ direction is compactified $\theta \sim \theta + 2\pi$ in the Euclidean geometry.

We would like to study the two-point correlation function
\begin{align}
\hat{G}(X_1,X_2,\theta) = \left \langle \chi(X_1,0) \chi(X_2,\theta) \right \rangle \ ,
\end{align}
which satisfies the equation
\begin{align}
\left(-\partial^2_{X_1} + U(X_1) - \partial^2_\theta \right) \hat{G}(X_1,X_2,\theta) = \delta(X_1 -X_2) \delta(\theta) \ . \label{grn}
\end{align}
The potential $U(X)$ is zero for the minimally coupled scalar, but it might get extra contribution from the coupling to the ``inflaton potential" that induces the phase transition.\footnote{We would like to thank the referee for pointing out that the minimally coupled scalar in two-dimension does not acquire any potential from the conformally flat metric.}

To solve \eqref{grn}, we first introduce the Green function for $\theta$ as
\begin{align}
G_k(\theta) = \sum_{n=-\infty}^{\infty} \frac{e^{in\theta}}{n^2 + k^2} \ ,
\end{align}
which satisfies
\begin{align}
(-\partial^2_\theta +k^2) G_k(\theta) = \delta(\theta) \ .
\end{align}
Then we use the completeness relation to write the delta function of $X$ as
\begin{align}
\delta(X_1-X_2) = \int_{- \infty}^{\infty} \frac{dk}{2\pi} u_k(X_1) u_k^*(X_2) +  \sum_i u_i(X_1) u_i^*(X_2) ,
\end{align}
where $u_k(X)$ are continuum spectrum while $u_i(X)$ are discrete spectrum \cite{Freivogel:2006xu}.
One can now solve the Green function in the form
\begin{align}
\hat{G}(X_1,X_2,\theta) = \int_{-\infty}^{\infty} \frac{dk}{2\pi} u_k(X_1) u_k^*(X_2)G_k(\theta) + \sum_iu_i(X_1) u_i(X_2) G_i(\theta) \ .
\end{align}

The continuum mode satisfies
\begin{align}
\left(-\partial^2_X + U(X)\right) u_k(X) = k^2 u_k(X) \ 
\end{align}
whose asymptotic form is specified by the boundary condition
\begin{align}
u_k(X) \to Te^{ikX} \  (\text{as} \ X \to -\infty) \ , \ \ u_k(X) \to e^{ikX} + R(X) e^{-ikX} \ (\text{as} \ X \to \infty) \ . 
\end{align}
As in \cite{Susskind:2007pv}\cite{Sekino:2009kv}, we only focus on the continuum contribution to the correlation function. Furthermore, the non-trivial part of the correlation function is encoded in the part proportional to the reflection coefficient $R(k)$. Then we obtain the relevant part of the two-point function as
\begin{align}
e^{X_1+X_2} \int \frac{dk}{2\pi} R(k) e^{ik(X_1+X_2)} \sum_n \frac{e^{in\theta}}{n^2 + k^2} \ .
\end{align}

After performing the analytic continuation and doing the integration over $k$, we obtain
\begin{align}
\langle \chi(T_1,0) \chi(T_2,Y) \rangle \sim \sum_{n=1} R(in) \frac{e^{-n|T_1+T_2|-n|Y|}}{n} \ . \label{supp}
\end{align}
The precise form of the reflection amplitude is not important, and the details depend on the coupling to the ``inflaton potential". In the thin wall approximation, the simple coupling to the wall gives the $\delta$-function potential for the scalar $\chi$.

In order to compute the boundary correlation function, we take the limit $T_1 \to \infty$ and $T_2 \to \infty$ to approach the conformal boundary in region I. Here we note that the conformal boundary is space-like unlike the FSSY in higher dimension, where it is light-like. Nevertheless, the most of the discussion there applies here: Up to the common boundary cutoff factor $e^{-nT}$, \eqref{supp} gives the massive powers of the boundary two-point functions on  the spatial boundary $Y$. As is proposed in \cite{Freivogel:2006xu}\cite{Sekino:2009kv} for the light-like boundary, we would reduce the theory in this extra dimension ($Y$ direction) to obtain two-dimensional lower theory. In this sense, the two-point correlation function \eqref{supp} can be seen as a superposition of the Kaluza-Klein reduced boundary correlation functions. In $(1+3)$ dimensional FSSY conjecture, the (Wick rotated) Kaluza-Klein momentum $n$ is related to the conformal dimension of the operator living on the two-sphere through this reduction. In our case, however, the two-dimensional reduced dual theory lives at a point, so there is no obvious notion of scaling dimension, nor correlation function. It is clear, however, that the ``field" in $(1+1)$ dimensional Liouville cosmology corresponds to a certain observable in the Kaluza-Klein reduced boundary theory. We will propose that the boundary theory is nothing but the old matrix model when $c<1$ in the next section.\footnote{Moreover, the conservation of the Kaluza-Klein momentum $n$ is only semiclassical approximation in the full quantum gravity as we will discuss in section 3.2.} Later, we will identify that the appearance of the ``massive" towers of the states in the zero-dimensional theory as the large $N$ matrix degrees of freedom.

The treatment of the cut-off factor $e^{-nT}$ needs care.
 As in the discussion in \cite{Sekino:2009kv}, the $T$ dependence in the ``massive towers" in the expansion \eqref{supp} is analogous to the cut-off dependence of the subleading contributions to the boundary correlation functions as studied in \cite{Gubser:1998bc}\cite{Witten:1998qj}\cite{Muck:1998rr} in the context of AdS/CFT correspondence. In the conventional AdS/CFT correspondence, the cut-off dependent subleading contributions are typically neglected because they are irrelevant in the formulation of the boundary conformal field theories, so it might seem reasonable to identify \eqref{supp} as a contribution from the single operator in the boundary but not from the summation over many operators.\footnote{However, in some situations, such cut-off dependence of the boundary correlation functions in the AdS/CFT correspondence may be physical and give us an interesting prediction \cite{Ho:2009zv}. When the FSSY conjecture is formulated with a manifest cut-off, the subleading part must be treated with care. Also, we note that in our matrix model living at a point, the notion of the Liouville factor they introduced to accommodate the extra time direction does not exist.} 
The complication here is that the factor does depend on the ``Kaluza-Klein" charge $n$, so as long as the Kaluza-Klein momentum is conserved, it makes sense to identify \eqref{supp} as a summation over the Kaluza-Klein operators. In relation to the cut-off, we point out that we do not treat the ``Liouville factor" of the boundary theory as dynamical variable, or rather we fix it to a particular value. This is related to the fact that in $(1+1)$ dimensional gravity, we also fix the gauge of the Weyl symmetry unlike higher dimensional gravity.

In principle, higher-point functions can be also computed from the analytic continuation from the Euclidean correlation functions. It is essential that the FSSY proposal is entirely based on the Euclidean formulation of the correlation function and its analytic continuation to the Lorentzian signature space-time. When the analytic continuation is well-defined, this gives us a definite way to choose quantum states in the time-dependent theory. In the next subsection, the intrinsic existence of the Euclidean theory plays a significant role in identifying a concrete boundary theory as the old matrix model when $c<1$.

\subsection{Matrix model description}
As in ordinary AdS/CFT correspondence, the holographic description of the eternal inflation in $(1+1)$ dimensional space-time is strong/weak duality. When the bulk gravity (i.e. Liouville field theory) is weakly coupled as $b\to 0$, where we need a large negative central charge for the matter sector, the dual holographic theory is supposed to have  a large degrees of freedom, and it will be strongly coupled.
On the other hand, when the Liouville central charge is small enough, the bulk gravitational theory is strongly coupled. In this parameter region, the boundary holographic theory is supposed  to be weakly coupled. In this subsection, we would like to propose that the boundary holographic theory is the old matrix model (in the genus zero limit for a single instanton geometry).

In the literatures, we have given two distinct interpretations for the matrix model describing the two-dimensional gravity coupled with minimal models. It was originally proposed as a dynamical discretization of the two-dimensional Euclidean world-sheet theory by using the 't Hooft expansion of the large $N$ matrix integral. There, the double scaling limit of the matrix integral in $1/N$ expansion can be thought as a generating function for the discretization of the Riemann surfaces. The other more recent interpretation is based on the space-time picture: we regard the matrix model as a large $N$ field theory of the localised D-brane (ZZ-brane) in the target space theory (\cite{McGreevy:2003kb}\cite{Douglas:2003up}\cite{Klebanov:2003wg}: see \cite{Nakayama:2004vk} for a complete list of reference). The claim is that the analogue of the gauge/gravity correspondence gives us an effective ``field theory" description of the gravitational theory from the boundary matrix model.

In this paper, we propose the third interpretation of the matrix model inspired by the FSSY conjecture. The matrix model describes the evolution of the $(1+1)$ dimensional Liouville gravity as the Kaluza-Klein reduction of the dual boundary theory living at the conformal infinity discussed in the previous subsection. The connection to the first interpretation is based on the analytic continuation. The philosophy of the FSSY conjecture is to compute the correlation function in the Euclidean signature, and through the discretization of the Riemann surface, the Euclidean computation can be done in the matrix model. The connection to the second interpretation is less obvious. We might imagine that the real time formulation of the string world-sheet may be described by a suitable analytic continuation of the ZZ-brane string field theory ($=$ matrix model) as in the Euclidean signature, where the world-sheet string theory is described by the ZZ-brane string field theory which is localized at the strongly coupled Liouville direction.

The matrix model is given by the simple matrix integration
\begin{align}
e^Z = \int DM \exp\left( - \mathrm{Tr} V(M) \right) \ . \label{matrixi}
\end{align}
The partition function (or rather free energy) $Z$ has a conventional $1/N$ expansion as
\begin{align}
Z = \sum_g N^{2-2g} Z_g(\kappa) \ ,
\end{align} 
where $\kappa$ is a symbolic way to denote coupling constants in the matrix integral (e.g. 't hooft coupling constant). The double scaling limit means that we take the critical limit $Z_g(\kappa) \sim (\kappa_c -\kappa)^{(2-\Gamma)(1-g)}$, where $\Gamma$ is the so called string succeptibility, and we rewite the partition function by using the double scaled parameter $\mu_r = N(\kappa- \kappa_c)^{(2-\Gamma)/2}$ as
\begin{align}
Z(\mu_r) = \sum_g \mu_r^{2-2g} f_g \ .
\end{align}
The higher critical point in the double scaling limit corresponds to introducing non-trivial matters in the gravity viewpoint. The critical points are connected via some integrable hierarchies (see \cite{Nakayama:2004vk} and references therein).

How do we see that the matrix model is a right way to describe the strongly coupled $c<1$ eternal inflation in $(1+1)$ dimension from the FSSY viewpoints? First of all, by construction we have one to one correspondence with the (local) field in the bulk theory and the (local) observables in the matrix model. The local field in the Euclidean Liouville gravity is well-known to be described by the matrix model correlation functions. In particular, turning on the relevant deformation in the gravitational theory corresponds to the ``flow" of the integrable ``times" in the matrix model (e.g. KP/KdV times in the KP/KdV hierarchy). Furthermore, more intuitively, one can compute the semi-classical higher point functions by applying the  FSSY procedure in the bulk. The higher-point interaction in the bulk Lagrangian that appears in the Landau-Ginzburg description $V(X) = X^{2(k-1)}$ corresponds to introduction of the non-trivial higher-point functions in the boundary theory. This is schematically consistent with the fact that we need a higher critical point to realize the matrix model corresponds to $k>3$. Also note that the number of boundary degrees of freedom (i.e. the size of the matrix $N$) is directly related to the (difference of the) cosmological constant $\mu$ of the Liouville gravity as in the higher dimensional FSSY proposal. 

At first sight, the matrix model contains infinite numbers of ``observables" at least obtained by the trace of higher powers of the matrix such as $\mathrm{Tr}M^n$, and as a consequence we need infinite numbers of ``operators" in the $(1+1)$ dimensional Liouville field theory to take account for them. We claim that the corresponding ``tower" appears in the bulk computation as a Kaluza-Klein tower \label{supp} in the imaginary Euclidean direction.
In addition, we also have ``non-local" operators (observables) in the Liouville field theory. Therefore, for instance, the $\mathrm{Tr}M^n$ for large $n$ describes the loop (or boundary) operators $\mathrm{Tr} e^{LM}$ in the two-dimensional Liouville gravity.
In the context of FSSY conjecture, it is not obvious how to interpret these non-local operators in the dual side, and the precise meaning of its Lorentzian continuation is not always clear. 
It would be interesting to further study this point. Intuitively, the loop-operators become a world-line operator, and it must describe a whole ``history" of the local observable.

 To establish the connection, it is imperative to recall that the FSSY conjecture is based on the computation  of correlation functions in the Euclidean signature and then perform the analytic continuation. This is in line with our matrix model dual proposal because as we know that the matrix model naturally describes the Liouville gravity in the Euclidean space. 


Note that the semi-classical field theory computation in section 4.1 for a simple matter field may not be directly applicable to our comparison with the matrix model computation. The Liouville gravity is strongly coupled with the Landau-Ginzburg field $X$, and the simple semiclassical treatment of the section 4.1 must be replaced by a full Liouville computation and a full matter CFT computation.
In other words, when we would like to perform the non-perturbative computation in section 4, we have to use the matrix model, so it is very natural to conclude that it is this matrix model that is the holographic dual of the two-dimensional Liouville cosmology.

As discussed in section 2.1, we recall that in the $\mu \to 0$ limit the boundary instanton is large, and the instanton amplitude can be seen as a sphere partition function with one-puncture. The scaling argument in the Liouville theory dictates that the one-point function should behave as $Z \sim \mu^{b^{-2}}$, and this is of course consistent with the matrix model computation proposed above. Furthermore, this Euclidean decay rate is identified with the $\mu \to 0$ limit of the tunnelling rate computed from the semi-classical instanton computation. The crucial assumption here, as we have stressed many times, is that the Lorentzian amplitude is obtained by the analytic continuation of the instanton amplitude, but this is the core of the FSSY conjecture.

As we have seen, The CDLZZ instanton and its Lorentzian continuation has a $U(1)$ isometry. It is interesting to ask what would be the corresponding symmetry in the matrix model. We point out that the matrix model as it is cannot and should not have such a manifest $U(1)$ symmetry. The point is that the matrix model provides the Euclidean partition function that will incorporate all the higher genus corrections. The CDLZZ instanton we have discussed just corresponds to the genus zero contribution of the matrix model. Obviously the multi-instantons (or multi-bubble formation with possible collisions in the Lorentzian interpretation) break the $U(1)$ isometry. In the matrix model picture, the $U(1)$ symmetry appears as an emergent symmetry associated with the restriction of the Feynmann diagram to the genus zero sphere (with puncture).\footnote{The matrix model has an obvious $U(N)$ ``gauge" symmetry. This has nothing to do with the space-time symmetry as in $U(N)$ gauge symmetry of the $\mathcal{N}=4$ super Yang-Milles theory.} 

It is nevertheless possible and illuminating to study the ``gauge unfixed" version of the matrix model partition function by introducing the ``Liouville factor" $\phi$ in the matrix model to imitate the ``boundary conformal invariance" of the higher dimensional FSSY conjecture:
\begin{align}
e^{Z'} = \int D'M D\phi \exp\left( - \mathrm{Tr} V(e^{\phi}M) \right) \ ,
\end{align}
where the measure $D'M = D(e^{\phi}M)$ is chosen so that it is invariant under the scaling transformation $\phi \to \phi + a$, and $M \to M e^{-a}$. The gauge unfixed version of the matrix model is useful to understand the cut-off dependence of the bulk theory through the boundary ``Liouville" factor \cite{Sekino:2009kv}. It is however immediate to see that the integration over the boundary ``Liouville" factor is trivial and it just yields the constant factor of ``volume of the scaling group": $\text{Vol}_{\phi} = \int_{-\infty}^{\infty} d\phi$ after the change of variable, and the remaining non-trivial part is just given by the gauge fixed partition function $e^{Z}$ in \eqref{matrixi}. As discussed in the last subsection, the triviality of the scaling factor in the boundary theory is due to the extra gauged ``Weyl (conformal) invariance" of the bulk gravity specific to the $(1+1)$ dimension.

With regard to the breaking of the $U(1)$ symmetry, let us discuss the multiple tunnelling and higher genus corrections to the eternal inflation in $(1+1)$ dimensional cosmology. In the Euclidean signature, we have to add all these multiple bubble solutions or higher genus instanton solutions to the Euclidean path integral. The Liouville path integral has a scaling behavior (for genus $g$):
\begin{align}
Z_g = f_g\mu^{(1-g)(1+b^{-2})}
\end{align}
and we will take multiple derivatives $\partial/\partial \mu$ to take into account the puncture contributions. This is precisely in accord with the classical Euclidean instanton computation in the Euclidean signature.

Let us now perform the Lorentzian continuation. Not all the Riemann surfaces can be analytically continued to the causal Lorentzian manifold. A simple counterexample is a torus: the Lorentzian continuation involves a closed time curve by definition. Others have natural Lorentzian interpretation such as multiple vacuum decay and the collisions of the bubbles and so on. For instance, the number of the puncture on the (multiple) Riemann surfaces correspond to number of bubbles in the geometry. The extra contribution $\mu^{b^{-2}}$ for an extra sphere bubble is precisely what we expect in the Lorentzian quantum tunnelling rate.

Finally, we remark that the way the non-trivial topology of the bubble contribute to the boundary theory is slightly different from the original FSSY conjecture in $(1+3)$ dimension \cite{Bousso:2008as}. There, the different topology of the bubble gives a boundary theory living on different topology. Here, the boundary theory is the matrix model independent of the boundary ``topology". The higher genus contributions are all encoded in the same matrix integral, and in this sense, our description is more ecological than their proposal.

\section{Further continuation --- Time-like Liouville theory}
In the previous sections, we have studied the dynamics of eternal inflation and quantum tunnelling in $(1+1)$ dimensional quantum gravity. The non-trivial instanton solution in the Liouville gravity has other interesting applications e.g. in $(1+3)$-dimensional, presumably more realistic, cosmology via the original FRW/CFT conjecture formulated in $(1+3)$ dimensional CDL universe. 

In this section, we study the analytic continuation of our results in the previous sections to the time-like Liouville theory. The time-like Liouville theory is ubiquitous besides the FRW/CFT conjecture: it appears everywhere when the matter central charge is large in two-dimensional gravity. For instance in \cite{Alishahiha:2004md}\cite{Dong:2010pm}, the time-like Liouville theory may appear in the context of the dS/dS duality. The quantum gravity effects in the (boundary) time-like Liouville field theory provides a non-trivial modification of the decay rate in the quantum cosmology in higher dimensions.
\subsection{Universe as holographic Liouville theory}
According to the original FSSY proposal \cite{Freivogel:2006xu} in $(1+3)$ dimensional cosmology, physics of the cosmic census taker is described by a boundary ``Liouville" theory living at the boundary of the Euclidean AdS$_3$ space that appears as spatial slices of the constant conformal time of the FRW universe:
\begin{eqnarray}
ds^2 = a(T)^2 (-dT^2 + ds^2_{\mathrm{EAdS}_3}) \ .
\end{eqnarray}
 Here, the Liouville theory appears as a holographic Wheeler-De Witt theory restricted to the boundary.\footnote{On another patch, it can be also represented as dS$_3$-CFT$_2$ correspondence, where the appearance of the Liouville gravity would be natural \cite{Strominger:2001pn}\cite{Klemm:2002ir}.} Unlike in the $(1+1)$ dimensional case studied in previous sections, the bulk gravity is dynamical, and not Weyl invariant, so the Liouville factor $T$ is also dynamical in the boundary.

In this setup, the (boundary) Liouville action is given by
\begin{eqnarray}
S = \frac{1}{4\pi} \int d^2x \left(\partial_a \phi \partial^a \phi + 4\pi \lambda e^{2b\phi} \right) \ ,
\end{eqnarray}
where the canonically normalized Liouville field $\phi = b^{-1}\varphi$ has been introduced (compare with \eqref{unl}). The Liouville field $\phi$ describes fluctuations of the boundary metric.

In order to apply this holographic description of the census taker to the realistic universe that contains many matter degrees of freedom, the matter sector should be included in the boundary theory as well. Since the central charge of the matter sector is large, the Liouville central charge $c_{\mathrm{Liouville}} = 1 + 6(b+b^{-1})^2$ should take a large negative value. This is only possible by setting imaginary Liouville exponent $b = i\beta$ with real $\beta$. 

The status of the Liouville theory with imaginary Liouville exponent, or time-like signature Liouville theory, which can be obtained after a further rotation of the field space $\phi \to iT$ to make the action real, is not as well-established as in the usual space-like case (see e.g. \cite{Gutperle:2003xf}\cite{Strominger:2003fn}\cite{Schomerus:2003vv}\cite{Fredenhagen:2003ut}\cite{Nakayama:2006gt}). The analysis below is based on the analytic continuation from the space-like Liouville theory and the intuition from the classical analysis. A different prescription of the analytic continuation sometimes leads to a different result, but we focus on the quantities that are not affected by such ambiguities.

A related point is the holographic interpretation of the Liouville cosmological constant $\lambda$ \cite{Susskind:2007pv}. It has nothing to do with the cosmological constant of the four-dimensional cosmological constant, but it is related to the time at which the census taker observes our universe:
\begin{eqnarray}
\lambda \sim e^{-2T} \ .
\end{eqnarray}
This is in accord with our intuition that the later the time, the more one can observe in our universe: the Liouville wall fades away as time goes.

\subsection{Fate of Unstable Census Taker}
In the previous subsection, we have reviewed the holographic description of the universe from the viewpoint of the census taker sitting at the future infinity. One implicit assumption there was that the universal is stable and the census taker can collect all the information in his causal patch that covers the whole FRW universe (and possibly its ancestors). However, in reality, our universe needs not to be eternally stable as long as it can accommodate some kinds of intelligence with sufficient long life-time. Indeed, the string landscape suggests, almost inevitably, that our universe is metastable \cite{Goheer:2002vf} and the natural end point would be a supersymmetric AdS vacuum after the decay of the current universe.

As a consequence, the cosmic census taker of our universe is to be treated as an approximate concept, and in particular, we would like to investigate the fate of our universe after its decay and the fate of the census taker from the holographic viewpoint. A natural candidate for such a description of the decay of the universe from the holographic Liouville filed theory would be to use the decay of the two-dimensional quantum Liouville gravity reviewed in section 2. Note that in any conformal field theory, in order to discuss the decay rate, we need a UV cut-off: otherwise the decay rate is infinite or zero (see \cite{Horowitz:2007pr} for a similar argument in the AdS/CFT correspondence). The introduction of the UV cut-off is actually an inevitable consequence of the approximate census taker in  the de-Sitter space \cite{Susskind:2007pv}.
Now, by giving it a holographic interpretation, we show how the decay of our four-dimensional universe is modified from the classical field theory result \cite{Kobzarev:1974cp}\cite{Coleman:1977py}\cite{Coleman:1980aw} by quantum gravity effects.\footnote{The notion of ``bounce instanton" here is slightly different from the boundary instanton introduced in \cite{Freivogel:2006xu}\cite{Susskind:2007pv} to discuss the bubble collisions in the eternal Minkowski space with eternal census taker. Technically, our bounce solution is a saddle point of the Euclidean action and drastically alters the physics both on the boundary and the bulk after the nucleation. See also \cite{deHaro:2006wy} for a similar discussion on the relation between the bulk instanton and the boundary instanton.}

We first assume that the decay happens after spending a long period in the current FRW universe, so the Liouville cosmological constant $\lambda \sim e^{-2T}$ is tiny and the metric of the Liouville sector is approximately flat. We also neglect the four-dimensional cosmological constant of the current universe to be within the FSSY conjecture.
Then, the decay process of the Liouville sector is approximated by the process studied in section 1. The energy difference $\mu$ (or more appropriately dimensionless ratio: $\mu/\sigma^2$) from the false vacuum of the boundary Liouville gravity to the true vacuum of the boundary Liouville gravity should be related to the vacuum decay rate of the current four-dimensional universe.\footnote{The interpretation of $\mu$ from the holographic perspective is interesting. Since the Liouville cosmological constant after the decay takes a negative value, the interpretation $-\mu = \lambda \sim e^{2T}$ demands that the conformal time $T$ is an imaginary number i.e. $T= \frac{i\pi}{2} + \frac{1}{2}\log|\mu|$. This kind of analytic continuation appears in the continuation from the Euclidean Coleman-De Luccia geometry \cite{Coleman:1980aw} to the Minkowski geometry.
The appearance of the imaginary time might be also related to the Hartle-Hawking approach to the emergence of the universe.} In the semiclassical limit in the both sides, the relation between the two  is obvious:
\begin{eqnarray}
P_{\mathrm{Liouville}} \sim \exp\left(-\frac{\pi\sigma^2}{\mu}\right) \sim \exp\left(-\frac{S(\Gamma)}{\hbar}\right) \sim P^{\mathrm{semi}}_{\mathrm{universe}} \ , \label{cld}
\end{eqnarray}
where the right hand side is the semiclassical vacuum decay rate of the current universe, and $S(\Gamma)$ is the bounce instanton action corresponding to the vacuum decay that could be computed from the effective four-dimensional field theory.

Since we know the quantum gravity corrections to the decay of the holographic universe encoded in the Liouville gravity, we can now understand the vacuum decaying rate of the four-dimensional universe including quantum gravity corrections by using the holographic relations. We begin with the decaying rate of the quantum Liouville gravity:
\begin{eqnarray}
P_{\mathrm{Liouville}} \sim \left(1+ \frac{\pi\sigma^2}{b^{-2}\mu}\right)^{-b^{-2}} \ . \label{ldecay}
\end{eqnarray}
In order to include matter contributions with large central charge $c_{\mathrm{matter}}$, we perform an analytic continuation $b \to i\beta$ to satisfy 
\begin{eqnarray}
26- c_{\mathrm{matter}} = 1 + 6(b+b^{-1})^2 = 1- 6(\beta-\beta^{-1})^2
\end{eqnarray}
as discussed in section 2. Under the analytic continuation, the decay rate now turns into
\begin{eqnarray}
P_{\mathrm{CT}} \sim \left(1-\frac{\pi\sigma^2}{\beta^{-2}\mu} \right)^{\beta^{-2}} \ . \label{key}
\end{eqnarray}
This is the main formula we propose in this section. It yields the decay rate of the current universe including the quantum gravity corrections from the boundary Liouville theory.

We first note that even after the analytic continuation, we can take a semiclassical limit: $\beta \to 0$ to study the semiclassical vacuum decay of the universe
\begin{eqnarray}
\lim_{\beta \to 0} P_{\mathrm{CT}} \sim \exp\left(-\frac{\pi\sigma^2}{\mu}\right) \ ,
\end{eqnarray}
which is consistent with the semiclassical decay rate of the current universe \eqref{cld}. A sizable difference from the semiclassical limit appears when $ \beta^{-2} \sim \frac{\pi \sigma^2}{\mu} = \frac{S(\Gamma)}{\hbar}$. We see that, for large  bounce instanton action, the decaying rate deviates from the semiclassical value, and interestingly, it is now more suppressed than the semiclassical result due to the quantum gravity corrections:
\begin{eqnarray}
P_{\mathrm{CT}} \sim \left(1-\frac{\pi\sigma^2}{\beta^{-2}\mu} \right)^{\beta^{-2}} < \exp\left(-\frac{\pi\sigma^2}{\mu}\right) \ .
\end{eqnarray}
As a result, the life-time of the metastable universe is enhanced by the quantum gravity corrections than naively expected from the bounce instanton computation in the bulk field theory.

In order to obtain this result, the analytic continuation of the Liouville exponent $b$ has been crucial. While the analytic continuation of the Liouville theory is not completely well established, we now show further evidence that our formula \eqref{key} is plausible by explaining, in particular, why we have obtained increased life-time instead of decreased life-time as in \eqref{zz}. We observe that at $\frac{\pi \sigma^2}{\mu} = \beta^{-2}$, the decay rate of the Liouville gravity, and hence, the decay rate of the universe with quantum gravity corrections vanishes. For larger value of $\frac{\pi \sigma^2}{\mu}$, even the proposed formula \eqref{key} does not make sense. If we interpret the situation in terms of the semiclassical decay rate computed from the four-dimensional field theory, the semiclassical decay rate has an interesting bound
\begin{eqnarray}
P^{\mathrm{semi}}_{\mathrm{universe}} \gsim \exp(-\beta^{-2}) \ . \label{dr}
\end{eqnarray}
As we will discuss in the next section, one may be able to understand the existence of the bound of the decay rate of the time-like Liouville theory from the maximum size of the possible Liouville bubble instanton in the {\it AdS space}. The reason why the AdS space is relevant here is that the classical equation of motion for the time-like Liouville theory is formally identical to that of the opposite cosmological constant with the positive kinetic term. As long as the classical extrema of the action is concerned, the value of the action can be computed by the instanton computation in the AdS space. Thus at the formal level, there exists a classical bounce instanton solution that corresponds to the analytically continued formula \eqref{key} in the time-like Liouville theory.
The action of the bounce instanton in the AdS space is computed in Appendix.

The bound says that the semiclassical decay rate of the current universe must be bounded from below to present any non-zero decay rate, and when the bound is violated, the corresponding decay channel can no longer exist after coupling to the quantum gravity. Note that the bound is characterized by the cosmological constant of the ancestor universe. The parameter $\beta^{-2}$ is related to the central charge of the matter sector $c_{\mathrm{matter}} \sim 6\beta^{-2}$ in the $\beta \to 0$ limit. On the other hand, according to the FSSY conjecture, the matter central charge is related to the entropy of the four-dimensional parent universe: $c_{\mathrm{matter}} \sim S_{\mathrm{universe}}$ \cite{Freivogel:2006xu}. Therefore, the bound on the decay rate \eqref{dr} shows that the daughter universe cannot admit the semiclassical decay rate over an extremely long time $T_{\mathrm{max}} \sim \exp(+S_{\mathrm{universe}})$ determined from the entropy of the parent universe. We note that $T_{\mathrm{max}}$ is nothing but the Poincar\'e recurrence time of the parent universe, so the classical (= field theory) decay channel in the current universe whose life-time is larger than the Poincar\'e recurrence time of the parent universe is forbidden by the quantum gravity effect.\footnote{Here we have neglected numerical factors for the entropy of the universe and the Poincar\'e recurrence time. It would be very interesting to discuss these factors and quantum corrections for finite $b$.} 

The approximation that the current universe has a zero cosmological constant relevant for the FSSY conjecture is violated at the Hubble scale set by the cosmological constant of the current universe. This implies that all the possible classical decay channels of the current universe whose decay rate is comparable with the Poincar\'e recurrence time of the parent universe are procrastinated at least until the Hubble scale of the current universe.  This quantum gravitational enhancement of the life-time of the universe, in particular, from the larger cosmological constant parent universe would be important in understanding of the evolution of the universe within the eternal inflation and to answer the measure problem therein. It would be also interesting to understand the effect of the breaking of the approximation of the zero-cosmological constant here because it would teach us the fate of the procrastinated decay till the (cosmological constant dominating) Hubble time of the current universe. It might suggest that as soon as the time scale of our universe is set by the cosmological constant, it would decay due to the procrastinated channels. 

\section{Discussion}
In this paper, we have proposed a concrete realization of the holographic description of the eternal inflation in $(1+1)$ dimensional Liouville gravity by applying the philosophy of the FRW/CFT correspondence. The core of the FSSY proposal is to understand the boundary correlation function from the Euclidean gravitational theory with suitable analytic continuation. It is this analytic continuation that made possible to identify the boundary theory as the old matrix model.

The validity of the analytic continuation still remains. In $(1+1)$ dimensional Lorentzian gravity, a simple discretization of the space-time path integral may not work, and the so-called causal dynamical triangulation has been developed for this purpose (see e.g. \cite{Ambjorn:2009rv} and references therein). So far, the causal dynamical triangulation approach has not been introduced the concept of quantum tunnelling, and it would be fascinating to see whether the quantum tunnelling can be incorporated in their formulation and compare the results with our holographic approach.\footnote{It has been recently shown that the causal triangulation approach reproduces the Liouville approach to the $(1+1)$ dimensional cosmology in \cite{Nakayama:1993we}, which seems encouraging.}

The relation between the Euclidean gravity and the Lorentzian gravity (and possible ``Wick rotation" or analytic continuation") would deserve further studies. It is a bit of shame that even in the $(0+1)$ dimensional gravity (i.e. world line theory of particle), the ``Wick rotation" is not well-established mathematically. Take a complicated Feynmann loop integral.  It is quite difficult to justify the momentum rotation $p_0 \to ip_4$ in {\it all} the (internal and external) momenta simultaneously. Cuts and poles will impede such procedures.\footnote{Of course, when we compute the beta function or deep Euclidean scattering, these subtleties do not matter in most cases.} We expect much to say about the validity of the analytic continuation in higher genus amplitudes and its interpretation in the FSSY conjecture. With this regards, the interpretation of the non-local observables in Euclidean Liouville gravity from the Lorentzian gravity would be of significance.

We have also investigated another analytic continuation to the time-like Liouville theory in this paper. The analytic continuation in the time-like Liouville theory is not straight-forward. For instance, naively speaking, if one studies the formal analytic continuation of the Liouville field theory for $c<1$, it is expected that the model is related to the so-called ``generalized minimal model". However, the Liouville correlation functions obtained simply put by $b\to i\beta$ does {\it not} give a minimal model correlation functions \cite{Zamolodchikov:2005fy}. This is due to the fact that the fundamental recursion relations in Liouville correlation functions cannot give a unique answer when $b$ is not a real parameter. Throughout the paper, we (as well as almost all the other literatures) have defined the time-like Liouville theory as such a formal analytic continuation. This is in accord with our spirit of the FSSY conjecture, but it would be important to establish the validity of this procedure.

When analytically continued to the time-like Liouville theory, our classical bounce instanton solution becomes a non-real solution, so we can never think of it as a classically realizable solution. It is, however, expected to be a relevant saddle point of the path integral in the WKB approximation. A similar non-real solution contributing to the path integral appears in various context such as high energy scattering amplitudes in string theory, or S-matrix computation in AdS/CFT correspondence \cite{Alday:2007hr} (see also \cite{Gross:1987kza} for the original flat Minkowski theory computation).

The formula we proposed as the analytic continuation of the space-like Liouville theory has very similar appearance with the formula for the decay rate of the Liouville theory from the negative cosmological constant to the zero cosmological constant (see Appendix A). In this situation, the bound for the classical action has a clear physical interpretation from the purely two-dimensional field theory viewpoint: the bounce instanton cannot be larger than the ambient space. It would be interesting to understand the proposed bound for the classical action in a similar manner. Note that  in the Lorentzian signature, again specific to the $(1+1)$ dimension, the change of the sign of the cosmological constant is equivalent to the change of the sign of the kinetic term at the level of the equation of motion.
 The coincidence of the decay rate might suggest a duality between the decay of the time-like Liouville theory and the decay of the space-like Liouville theory with cosmological constant reversed.\footnote{At the minisuperspace approximation, the connection between the two has been pointed out in \cite{Gutperle:2003xf}. See also \cite{McGreevy:2005ci}\cite{Nakayama:2006gt}.}

The relation between the bound on the classical instanton action in our universe and the Poincar\'e recurrence time of our ancestor universe is non-trivial. This non-locality in the quantum gravity is the manifestation of the FSSY conjecture that connects the physics of our universe and the physics of our ancestor universe. The discussion was restricted to the single decay, and it would be interesting to see what happens when the universe admits a multiple decay, and the subsequent bubbles collision.




\section*{Acknowledgements}
The research is supported in part by NSF grant PHY-0555662 and the UC Berkeley Center for Theoretical Physics. The author would like to thank B.~Freivogel for stimulating discussions.

\appendix
\section{Decay of Liouville gravity with generic cosmological constant}
In \cite{Zamolodchikov:2006xs}, they studied the decay rate of the two-dimensional quantum gravity from zero cosmological constant to the negative cosmological constant. Here we would like to investigate a similar instanton correction but from for the generic cases between vacua with negative cosmological constants.\footnote{At the end of this appendix, we discuss the more generic combinations of positive and negative cosmological constants.}

The topology of the Euclidean instanton is a sphere. The Liouville action outside the bubble is given by
\begin{eqnarray}
S^{\mathrm{OUT}}_{\mathrm{Liouville}} = \int_{r > |z|} d^2 z \left(\frac{1}{4\pi b^2}\partial^a\varphi \partial_a \varphi - \mu e^{2\varphi} \right)\ ,
\end{eqnarray}
and inside the bubble, it is given by 
\begin{eqnarray}
S^{\mathrm{IN}}_{\mathrm{Liouville}} = \int_{r < |z|} d^2 z \left(\frac{1}{4\pi b^2}\partial^a\varphi \partial_a \varphi - \bar{\mu} e^{2\varphi} \right)\ ,
\end{eqnarray} 
where $r$ is the radius of the bubble. We set $\bar{\mu} > \mu >0$. 

Outside the bubble, the Liouville equation of motion (with the ``wrong" cosmological constant)
\begin{eqnarray}
\partial \bar{\partial} \varphi = -\pi \mu b^2 e^{2\varphi}
\end{eqnarray}
can be solved as
\begin{eqnarray}
e^{2\varphi} = \frac{4R^2 a^2}{(1+a^2z\bar{z})^2} \ , \ \ \ \ R^2 = \frac{1}{4\pi\mu b^2} \ . \label{poincare1}
\end{eqnarray}
Similarly, inside the bubble,  the Liouville equation of motion
\begin{eqnarray}
\partial \bar{\partial} \varphi = -\pi \bar{\mu} b^2 e^{2\varphi}
\end{eqnarray}
can be solved as
\begin{eqnarray}
e^{2\varphi} = \frac{4\bar{R}^2 a^2}{(1+\bar{a}^2z\bar{z})^2} \ , \ \ \ \ \bar{R}^2 = \frac{1}{4\pi\bar{\mu} b^2} \ . \label{poincare2}
\end{eqnarray} 

The parameters $a$, $\bar{a}$ and $r$ are mutually related through the continuity condition for the Liouville field. By using the scaling symmetry, one can impose the normalization condition $\phi = 0$ at $|z|=r$. 
The parameters $a$ and $\bar{a}$ are determined from the continuity condition at $|z| = r$ as
\begin{align}
2Ra &= 1 + a^2 r^2 \cr
a &= \frac{R \pm \sqrt{R^2-r^2}}{r^2} \ ,
\end{align}
and
\begin{align}
2\bar{R}\bar{a} &= 1 + \bar{a}^2 r^2 \cr
\bar{a} &= \frac{\bar{R} \pm \sqrt{\bar{R}^2-r^2}}{r^2} \ ,
\end{align}
Following the convention in \cite{Zamolodchikov:2006xs}, we introduce $t= ar$ (and $\bar{t} = \bar{a} r $) as a uniformizing parameter.

We can compute the difference of the one-instanton action $S_{\mathrm{inst}}$ and the zero-instanton action $S_0$ as
\begin{eqnarray}
S_{\mathrm{inst}} - S_0 = -\int_{r<|z|} d^2z \left( \frac{1}{4\pi^2 b^2} \partial_a \varphi \partial^a \varphi - \mu e^{2\varphi} \right) + \int_{r<|z|} d^2z \left( \frac{1}{4\pi^2 b^2} \partial_a \bar{\varphi} \partial^a \bar{\varphi} - \bar{\mu} e^{2\bar{\varphi}} \right) \ ,
\end{eqnarray}
where $\varphi$ is given by \eqref{poincare1} and $\bar{\varphi}$ is given by \eqref{poincare2}.
We can evaluate the integral as
\begin{eqnarray}
\int_{r<|z|} d^2z \frac{1}{4\pi^2 b^2} \partial_a \varphi \partial^a \varphi  = b^{-2}\left(\log(1+t^2) - \frac{t^2}{1+t^2} \right) 
\end{eqnarray}
and
\begin{eqnarray}
\int_{r<|z|} d^2z  \mu e^{2\varphi} = 4\pi\mu R^2 \frac{t^2}{1+t^2} \  
\end{eqnarray}
and similarly for $\bar{\varphi}$. 

By adding the surface tension term $2\pi \sigma r$, the total action becomes
\begin{align}
&S_{\mathrm{inst}}-S_0 \cr
=& -b^{-2}\left(\log(1+t^2) - \frac{t^2}{1+t^2} - x^2 (1+t^2) - 2 s x\right. ) \cr
&- \left. \log(1+\bar{t}^2)
+ \frac{\bar{t}^2}{1+\bar{t}^2} +\bar{x}^2(1+\bar{t}^2) + 2\bar{s}\bar{x} \right) \ ,
\end{align}
where $2x = r/R$, $2\bar{x} = r/\bar{R}$ and $\frac{s}{b^2 R} - \frac{\bar{s}}{b^2 \bar{R}} = 2\pi \sigma$. The action can be extremized when $t= s$, $\bar{t} = \bar{s}$, $x = \frac{s}{1+s^2}$ and $\bar{x} = \frac{\bar{s}}{1+\bar{s}^2}$, which yields the on-shell action:
\begin{align}
-\bar{S}_{\text{inst}} + \bar{S}_0 = b^{-2}\log\left(\frac{1+s^2}{1+\bar{s}^2}\right) \  
\end{align}
with 
\begin{align}
s &= \frac{-\bar{R}^2 + R^2 + \tilde{\sigma}^2 - \sqrt{4 R^2 \tilde{\sigma}^2 +(\bar{R}^2 - R^2 + \tilde{\sigma}^2)^2}}{2 \bar{R} \tilde{\sigma}} \cr
\bar{s} &= \frac{-\bar{R}^2 + R^2 - \tilde{\sigma}^2 - \sqrt{4 R^2 \tilde{\sigma}^2 +(\bar{R}^2 - R^2 + \tilde{\sigma}^2)^2}}{2 {R} \tilde{\sigma}} \ ,
\end{align}
where $\tilde{\sigma} = 2\pi b^2 R \bar{R} \sigma $ is a rescaled domain wall tension. Note that the formula reduces to \eqref{zz} by taking ${R} \to \infty$.

Analogous computation gives a decay of the Liouville gravity from the negative cosmological constant (to the positive cosmological constant).
Without going into the detailed derivation, which is straightforward but not illuminating, we present the final decaying rate formula of the two-dimensional gravity from cosmological constant $\mu$ to $-\bar{\mu}$ under the assumption that the change of the central charge is negligible before and after the decay. The result is 
\begin{eqnarray}
e^{-\bar{S}_{\mathrm{inst}} + \bar{S}_0} = (1-s^2)^{b^{-2}} (1+\bar{s}^2)^{-b^{-2}} \ ,
\end{eqnarray}
where 
\begin{eqnarray}
s &=& \frac{\bar{R}^2+R^2 +\tilde{\sigma}^2+ \sqrt{4R^2\tilde{\sigma}^2 + (\bar{R}^2 + R^2 -\tilde{\sigma}^2)^2}}{2\bar{R}\tilde{\sigma}} \cr 
\bar{s} &=& \frac{\bar{R}^2+R^2 -\tilde{\sigma}^2 + \sqrt{4R^2\tilde{\sigma}^2 + (\bar{R}^2 + R^2 - \tilde{\sigma}^2)^2}}{2R\tilde{\sigma}} \ .
\end{eqnarray}
Here we have introduced $R^2 = \frac{1}{4\pi\mu b^2}$ , $\bar{R}^2 = \frac{1}{4\pi \bar{\mu} b^2}$ and the rescaled surface tension $\tilde{\sigma}$. The formula can be formally regarded as an analytical continuation $\mu \to -\mu$ in the previous formula. In particular, when $\bar{R} \to \infty$, we obtain the decay rate $P \sim (1-s)^{b^{-2}}$, which we have cited in section 5.
An intuitive reason why $s$ cannot be larger than $1$ is that the size of the instanton cannot be larger than the AdS radius. Further analytic continuation $\bar{\mu} \to -\bar{\mu}$ gives the decay of the Liouville gravity between vacua with positive cosmological constants.

\end{document}